%% file: main.tex
\documentclass[english,aps,prstper,reprint,showpacs,titlepage,longbibliography]{revtex4-2}   

\usepackage[T1]{fontenc}	
\usepackage[latin9]{inputenc}	
\usepackage{geometry}		
\usepackage{graphicx}
\usepackage[above,below]{placeins}	
\usepackage{times}
\usepackage{makecell}

\usepackage{hyperref}  
\hypersetup{colorlinks=true,urlcolor=blue,citecolor=blue,linkcolor=blue}   
\usepackage{enumitem}          
\setlist{nosep}                 
\usepackage{multirow}
\usepackage{xcolor, soul}
\sethlcolor{black}

\begin{document}

\begin{titlepage}

\title{Industry Insights into Quantum Knowledge Needed for the Quantum Information Science and Engineering Workforce}

  \author{A.R. Pi\~{n}a\textsuperscript{1}}
  \author{Shams El-Adawy\textsuperscript{2,3}}
  \author{H.~J. Lewandowski\textsuperscript{2,3}}
  \author{Benjamin M. Zwickl\textsuperscript{1}}
  \affiliation{\textsuperscript{1}School of Physics and Astronomy, Rochester Institute of Technology, Rochester, NY, 14623} 
  \affiliation{\textsuperscript{2}JILA, National Institute of Standards and Technology and the University of Colorado, Boulder, Colorado, USA} 
  \affiliation{\textsuperscript{3}Department of Physics, University of Colorado, Boulder, Colorado USA}


  \begin{abstract}
   Quantum Information Science and Engineering (QISE) education and workforce development are  top priorities at the national level in the US. This has included a push for academia to support the development of programs that will prepare students to enter the QISE workforce. As the field of QISE has grown rapidly in academia and industry, there is a need to better understand what quantum knowledge is needed for students to be ready for the workforce. We present preliminary findings on the level of quantum expertise and the specific quantum knowledge utilized across different roles, and in the execution of specific tasks in the QISE industry. Qualitative analysis of semi-structured interviews with industry professionals elucidates these aspects of the vital work functions related to the ongoing development of quantum technologies in industry. This work will provide insights into QISE curriculum development and changes needed to better support students transitioning into this growing industry. \\
 \clearpage
\end{abstract}

\maketitle
\end{titlepage}

\section{Introduction}
Quantum information science and engineering (QISE) has developed rapidly in recent years. 
While the field has strong roots in physics and the academy, it is now highly interdisciplinary \cite{pina2025landscape, meyer_interdisc_classrom} and has a growing industry \cite{NQI, Fox2020_workforce}. 
However, the diversity in focus and approaches to the different areas of quantum technology (computing, networking, and sensing), as well as the rapidly evolving nature of the field, have made it difficult to develop a thorough understanding of the QISE careers available to STEM students and what knowledge and skills they would need in order to be prepared for them \cite{Fox2020_workforce, hughes2022, Oliver_qforge_2025, hasanovic2022}.
The QISE industry also has many stakeholders: educators who would like to ensure their students have been appropriately prepared to enter the workforce \cite{goorney_2025, quantumeducators, meyer_intro}, companies that would like to hire qualified and productive candidates \cite{Fox2020_workforce}, and government agencies that want to facilitate a robust QISE education ecosystem \cite{NQI, CHIPS, NSP}. 

Quantum and QISE education has also been growing in interest among various academic disciplines.
There have been large-scale efforts to assess the landscape of quantum and QISE education \cite{Meyer2024, pina2025landscape, Cervantes2021}, as well as more detailed efforts to understand the allocation of classroom time across different topics \cite{buzzell_quantum_2024, meyer_intro}. 
Although these studies are providing insight into the present state of QISE education at an unprecedented scale, it is unclear the extent to which these educational efforts are preparing students for careers in QISE industry, partly due to the  evolving needs of the industry.

In an effort to characterize the needs of the industry in 2019, \citet{Fox2020_workforce} interviewed 26 individuals  from  21 distinct companies in the QISE industry. 
Analysis of these data elucidated the knowledge and skills  valued by employers, however, there is the potential for new findings or changes in industry priorities given both the growth of the industry and evolution of the technology since the work of \citet{Fox2020_workforce}.
Other recent studies have focused on primarily survey data, which does not allow for the same depth of analysis as interview data\cite[e.g.,][]{hughes2022}.
Interview data are especially valuable for curriculum design because they reveal how the knowledge, skills, and abilities (KSAs) are applied in practice, which can help determine the appropriate depth and focus of instruction.

Although there are various categories of KSAs necessary for one to be successful in the QISE industry, here we present the subset most directly related to quantum. 
Across a sample of 42 QISE job roles in our dataset, we will address the following research questions: 1) What level of quantum expertise is necessary for different job roles and what are some of the quantum related KSAs necessary for those roles? 2) What are the quantum KSAs that are required for specific tasks in the QISE industry?

\section{Methods}
In order to gain insight into the KSAs required for different roles and tasks in the QISE industry, we conducted 27 semi-structured interviews that lasted 30~$\textendash$~ 60 minutes with individuals currently employed at one of 16 different QISE companies. 
Recruitment included multiple strategies: contacting individuals at companies involved with the Quantum Economic Development Consortium (QED-C)  \cite{qedc_homepage}, leveraging our professional networks across industry and academia, and snowball sampling \cite{parker2019snowball}. 
Our participants came from a wide range of company types, including quantum sensing, quantum networking and communication, quantum hardware, quantum algorithms and software, enabling technologies, and consultants. 

We conducted the interviews with two different protocols: one protocol was aimed at individuals in managerial roles and the other was aimed at employees. 
Portions of the protocols were written to address specific components of the O*NET Content Model \cite{onetContentModel}, which is a framework intended to identify and categorize six components of an occupation: worker characteristics, worker requirements, experience requirements, occupational requirements, occupation-specific information, and workforce characteristics. The O*NET framework also includes definitions of KSAs and tasks that were used for our analysis, however, we did not address ``abilities'' explicitly in this work. As we showcase in the rest of this section, the O*NET framework guided the structure of our interview protocol, and analysis of roles in the QISE industry.

We asked managers to discuss up to five different roles related to QISE at their company after providing them with a definition of a quantum-related role, which was ``A job directly related to or supporting the theory, implementation, or development of quantum information science and technology.'' We use the terms ``job'' and ``role'' interchangeably.
For each role, we asked participants for a general description of the position (occupation-specific information), the educational level and discipline(s) needed (worker requirements), prior work experience required (experience requirements), and a series of questions about the KSAs required for the role (occupation specific information \& occupational requirements). We divided the questions about KSAs into three primary categories: quantum specific (e.g., understanding spin-orbit coupling or laser cooling), general scientific (e.g., coding or troubleshooting), and professional (e.g., communication and collaboration). Interviews with employees were structured similarly, but instead of asking participants to discuss different roles at the company, we asked them to discuss up to five tasks that they consider critical to their own role at the company.  For each task, we asked employees to discuss the same three sets of KSAs (quantum, scientific, and professional) utilized for the task.

\begin{figure*}[htb]
    \centering
    \fbox{\includegraphics[width=.8\linewidth]{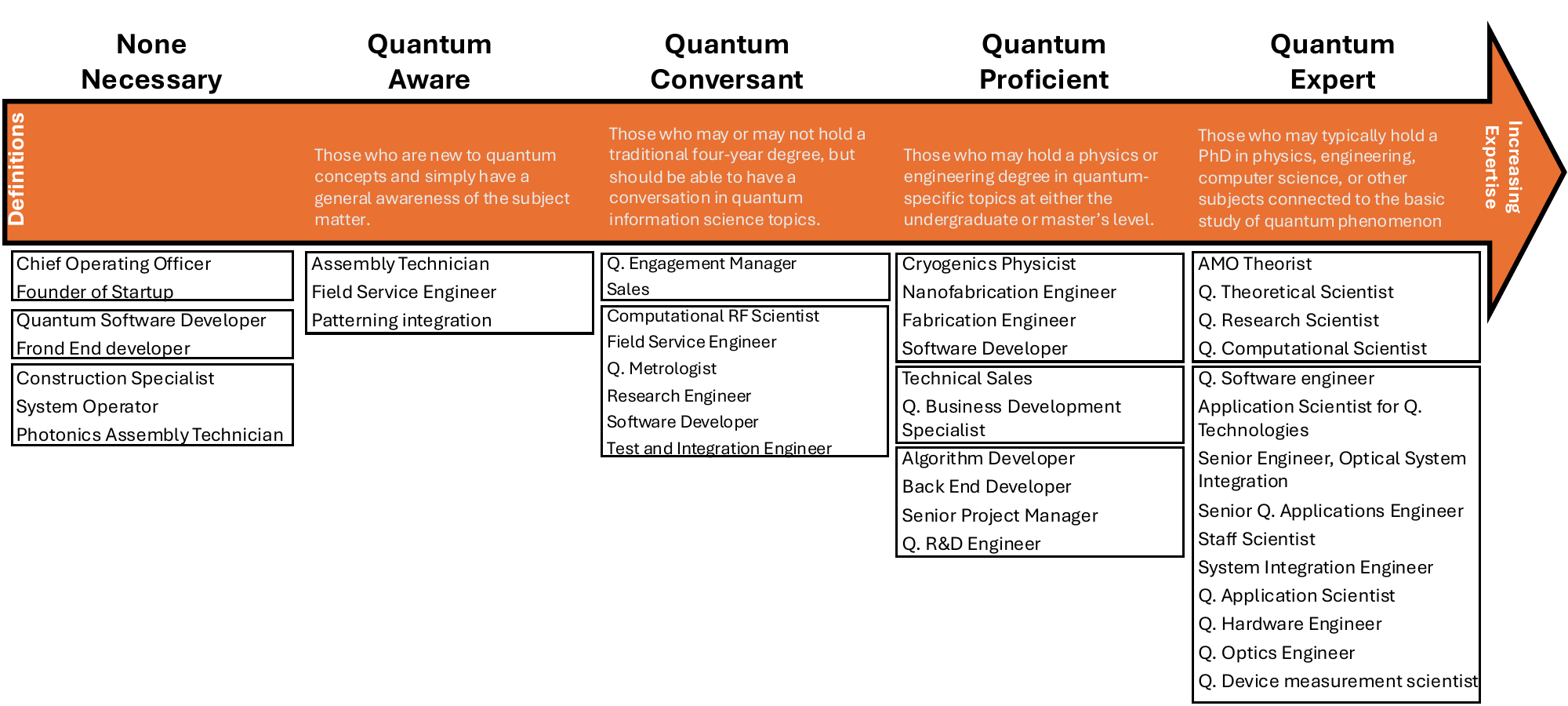}}
    \caption{Job roles in the dataset sorted by the required level of quantum expertise, and boxed by sub-themes discussed in our analysis. Abbreviated definitions of the five different levels of quantum expertise are given within the arrow, which indicates the direction of increasing level of quantum expertise. `Quantum' has been shortened to `Q.'}
    \label{fig:roles}
\end{figure*}

A key feature in both of our interview protocols is that we asked managers and employees to select the requisite level of quantum expertise for each role discussed. This classification included the following levels: quantum aware, quantum conversant, quantum proficient, or quantum expert, which were derived from a \citet{cubit2024roadmap} report on the quantum workforce. Abbreviated definitions for these levels of quantum expertise are provided in Fig. \ref{fig:roles} in Sec.~\ref{sec:results} and were also provided to all participants during the interview.

To answer our research questions, we focused our preliminary analysis on interview participants' responses to the quantum expertise level, as well as the examples and descriptions they provided about the type of quantum KSAs required for different roles and tasks. We began by collecting interview participants' responses regarding quantum expertise level.  Then, within each expertise level, we identified sub-themes that reflected distinct ways of engaging with quantum knowledge in practice. Afterwards, we organized the sub-themes into broader categories of roles and task types, which allowed us to get a deeper understanding of how quantum knowledge and skill requirements varied across the expertise spectrum. 

There are a few limitations to our data analysis process that we may be able to address with further data collection and analysis. 
First, for the scope of this paper, we focus only on quantum KSAs and exclude participants' discussion of scientific and professional KSAs.  Although this deliberate choice allows us to focus on how much quantum is needed to work in various roles in the QISE industry, it omits some quantum knowledge that may have been discussed in participants' answers about scientific KSAs. 
Second, in our analysis, we noticed that the level of quantum expertise of the participants also leaned more toward the quantum expert than the quantum aware level (see Table \ref{tab:tasks}). This is in part due to this initial sample of interviews we collected, as we sought to first collect data on roles most closely related to quantum. Nonetheless, this means there is the potential that we are missing some relevant KSAs required for tasks done by employees who have a lower level of quantum expertise. 

\section{Results}
\label{sec:results}

\subsection{QISE Job Roles}
Across the 27 interviews, 42 different roles in the QISE industry were discussed. 
We organized the roles primarily according to their associated required levels of quantum expertise (quantum aware, quantum conversant, quantum proficient, or quantum expert), which are defined in Fig. \ref{fig:roles}. 
However, an additional category was added to account for roles that did not require any quantum expertise. For roles requiring quantum expertise,  we describe the quantum KSAs required.
For the roles that do not involve any quantum-specific expertise, we instead summarize their primary responsibilities and explain their importance within QISE companies.

\subparagraph{No quantum requirements:}
There were a variety of roles reported for which there was no requisite quantum expertise. 
For individuals in these roles, it could be helpful to know something about the underlying theory, but it is not necessary to begin working in the roles. 
Within these seven roles, we identified three sub-themes: business roles, software roles, and hands-on roles. 
An example of one such business role is the chief operating officer whose primary responsibilities include overseeing business development, financial projection, proposal writing, and setting the company road map.
A software role example, such as a front end developer, builds chemistry modeling interfaces and generates visualizations and dashboards for different end-users of quantum technology. 
The hands-on roles often required some knowledge of optics and the ability to work in an optics lab. 

\subparagraph{Quantum aware:}
Roles requiring some level of quantum expertise, but not deep understanding involve practical, hands-on skills, such as fiber splicing, building optical setups, or integrating conventional manufacturing techniques with those used in quantum technologies, and require an awareness of how their work contributes to the broader goals of a specific quantum technology.

\subparagraph{Quantum conversant:}
Roles that require a conversational level of quantum expertise, fall into two sub-themes: public-facing roles and technical roles. 
The public-facing roles, such as sales and engagement, involve direct interaction with clients to understand their needs and communicate how the company's quantum products or services can address specific challenges. 
Individuals in these roles must be able to synthesize and convey key ideas about quantum technologies in accessible ways. 
While they do not require deep technical expertise, a strong background in physics or a closely related field involving quantum concepts is essential. 
The technical roles requiring this level of quantum knowledge include computational radio frequency scientists and software developers. 
These individuals must understand how their work connects to that of colleagues in quantum-proficient or quantum-expert roles. 
They also need a foundational understanding of quantum circuits and device physics, even though they are not directly responsible for implementing quantum systems.

\subparagraph{Quantum proficient:}

Among roles requiring quantum proficiency, some require advanced knowledge in a specific technical domain, such as nanofabrication, cryogenics, or fabrication, combined with a working knowledge of quantum mechanics. 
For example, a nanofabrication engineer may need to understand quantum emitters; a cryogenics physicist must be familiar with cryogenic testing of quantum systems and sensors; and a fabrication engineer (distinct from a nanofabrication engineer) should be able to design and troubleshoot magneto-optical traps.

Another sub-theme includes quantum-proficient software developers, whom are expected to understand quantum circuits in depth, including their mathematical representations and operations. 
They must also be capable of interpreting the results of theoretical quantum computations and possess insight into current challenges and emerging trends in quantum computing.
Technical sales and quantum business development specialists, however, are less focused on research or manufacturing, but still require strong quantum knowledge. 
Both roles demand an understanding of QISE theory and practice, and how quantum technologies can be applied across industries. 
For instance, technical sales professionals must have an understanding of the inner workings of quantum computers, including sources of noise, to make informed recommendations to customers. 
Similarly, business development specialists must identify the gap between a potential customer's current needs and the advantages offered by quantum technologies.

The remaining roles requiring quantum proficiency require specialized quantum knowledge and skills.
Individuals in these positions must be able to design, build, and work with various quantum systems, qubits, and gates. 
These roles require a deep understanding of both theory and experimental implementation, including topics such as open quantum systems, quantum error correction, (de)coherence, and cryptography.

\subparagraph{Quantum expert:}
AMO theorist, quantum theoretical scientist, quantum research scientist, quantum computational scientist, and quantum software engineer are roles for quantum experts that all require primarily theoretical knowledge, including an understanding of open system theory, process matrices, noise modeling, quantum algorithms, and error correction. 
Other roles, such as applications and engineering roles (see Fig. \ref{fig:roles}) require highly sophisticated hands-on skills, including experience working with various qubit platforms, quantum optics experiments, and device characterization. 
These hands-on roles also require a deep theoretical understanding of their experimental work, which stands in contrast to some of the hands-on quantum proficient or conversant roles that may require lab skills with less theoretical understanding. 

\subsection{Quantum Knowledge and Skills for Specific Tasks}
Interviews with employees in the QISE industry revealed the quantum KSAs central to various tasks associated with roles. 
We identified three categories of tasks defined by their quantum KSAs which are summarized in Table \ref{tab:tasks}.
The first category of tasks involves primarily hands-on experimental work. 
These tasks require advanced lab skills relevant to quantum systems, along with an understanding of the underlying theoretical concepts.
Tasks in the second category emphasize theoretical quantum and QISE knowledge. 
They also require an understanding of experimental techniques and practices, but do not necessarily require the explicit use of experimental skills. 

The final category of tasks require only theoretical knowledge of quantum or QISE. 
These tasks are largely centered on teaching, publishing, or engaging with published literature related to QISE. 
Notably, the people in roles executing these tasks are not explicitly theory roles. 
We do not yet have task-level data for QISE theorists in industry and anticipate that we will find tasks more central to the development and implementations of novel technologies as we continue to collect more data. 
The theory only tasks here do, however, represent the utility of QISE theory to some of the more hands-on roles such as application scientists.
If we examine the quantum expertise of the roles associated with tasks (rightmost column in Table \ref{tab:tasks}), we notice that there are multiple quantum proficiency levels (e.g., none, aware, conversant, proficient, expert) represented in two categories (Experimental, Theory and Knowledge of Experiment), though a more comprehensive reporting could reveal an even wider range of proficiency levels. This suggests that different levels of are important for all three categories, though the Theory Only category is largely experts.

\begin{table*}[htb]
\caption{Examples of critical tasks in the QISE industry and the quantum knowledge and skills required for those tasks as reported by interview participants. These tasks have been categorized by the nature of the quantum knowledge and skills required for them.}    \label{tab:tasks}
    \centering
    \begin{tabular}{p{.8in}|p{1.6in}|p{2.9in}|p{1.5in}}
           \hline \hline
      \multicolumn{1}{c|}{\textbf{Category}}& \multicolumn{1}{c|}{\textbf{Task}} & \multicolumn{1}{c|}{\textbf{QM Knowledge and Skills}}& \multicolumn{1}{c}{\textbf{Role}}  \\
       \hline 
       
       \multicolumn{1}{p{.75in}|}{\multirow{3}*{\textbf{Experimental}}} & Test of quantum systems and subsystems  & Understanding of, and experience working with, integrated systems that support trapped ions&Test and Integration Engineer (Quantum Conversant)\\\cline{2-4}

        & Demonstration and applications engineering & High literacy of quantum knowledge and experimental skills including the ability to implement experimental setups form literature&Quantum R\&D Engineer (Quantum Proficient)\\ \cline{2-4} 

        & Quality testing & Operating magneto-optical traps and interpreting results of system measurements&Fabrication Engineer (Quantum Proficient)\\
\hline
       \multicolumn{1}{p{.75in}|}{\multirow{3}*{\textbf{\makecell{Theory and \\Knowledge of \\Experiment}}}}& Customer Communication about research projects&Knowledge of different qubit types, measurement setups, and wiring required for different experiments.&Application Scientist for Quantum Technologies (Quantum Expert) \\\cline{2-4} 

      & Technical sales support & Understanding of specific applications of quantum technologies and a broad understanding of quantum technology implementation. &Senior Product Manager (Quantum Proficient) \\\cline{2-4} 

        & Software package development & Theoretical understanding of quantum circuits including the mathematical representations of their operations and different measurement processes.&Software Developer (Quantum Proficient)\\
\hline
       \multicolumn{1}{p{.75in}|}{\multirow{3}*{\textbf{Theory Only}}}  & Reading and Writing Research Papers & Advanced theoretical understanding of quantum and other areas of physics& Senior Quantum Applications Engineer (Quantum Expert)\\\cline{2-4} 

       & Staying up-to-date in QISE advancements &  Some understanding of most qubit hardware platforms, quantum gates, error correction, and quantum Technology Implementation & Application Scientist for Quantum Technologies (Quantum Expert)\\\cline{2-4} 

        & Train engineers on QISE topics & PhD level knowledge of quantum and quantum computing. &Senior Engineer Optical System Integration (Quantum Expert)\\
                   \hline \hline

    \end{tabular}

\end{table*}

\section{Discussion}
We presented some preliminary findings on the level of quantum expertise required for different roles in the quantum industry and the quantum knowledge and skills associated with specific tasks for a subset of those roles. 
Within the different levels of quantum expertise, we identified groups of roles that fell into various sub-themes that reflected distinct ways of engaging with quantum knowledge in practice.
Among roles requiring no quantum expertise, the sub-themes revolved around business, software development, and hands-on work. 
At the levels quantum aware and quantum conversant, roles in our sample tend to require an understanding of how an individual's work fits into the broader scheme of quantum technology and the ability to interface with people at higher levels of quantum expertise. 
As the level of quantum expertise increases to quantum proficient and quantum expert, so does the degree of specialization of the roles. 
This is demonstrated by the differences between hands-on and computational/theory roles requiring higher levels of quantum expertise. Among tasks discussed by QISE employees we identified three categories of required quantum KSAs: experimental,theory and knowledge of experiment, and theory only.   

Part of the utility in the role-level data and categorizations for students and educators is in building general awareness of career options. 
There is evidence that students are largely unaware of what roles exist in the QISE industry and what is required for them \cite{Oliver_qforge_2025} and elucidating the variety of roles in the industry could help address this.
The task-level data is specific enough that it could have implications for instruction. 
Previous research has shown that QISE educators need a more thorough understanding of KSAs utilized in industry so they can develop more relevant course goals and learning objectives \cite{quantumeducators}. 
In particular our future work will seek to elucidate any relationships that exist between the necessary KSAs and different company types in the QISE industry. 
As we continue to collect data on more roles and tasks, we anticipate emergent themes that could suggest new units, courses, or programs in QISE education. 
Taken holistically, these data suggest that there could be a more detailed categorization of quantum KSAs than the aware-expert spectrum.

\acknowledgments
This material is based on work supported by the Army Research Office under Award Number: W911NF-24-1-0132 and the National Science Foundation under Grant Nos. PHY-2333073 and PHY-2333074. \cite{asee}

\clearpage

\input{main.bbl}

\end{document}

%% file: main.bbl
%